\documentclass[aps,nofootinbib,showpacs,showkeys,preprint]{revtex4}

\usepackage{epsf,epsfig,subfigure, axodraw, graphicx,amsmath,amssymb}
\usepackage{color}

\newcommand{\dis}[1]{\begin{equation}\begin{split}#1\end{split}\end{equation}}
\newcommand{\be}{\begin{equation}}
\newcommand{\ee}{\end{equation}}
\def\bea{\begin{eqnarray}}
\def\eea{\end{eqnarray}}

\def\ptmiss{\not\!\!{p_T}}

\newcommand{\eq}[1]{Eq.~(\ref{#1})}

\begin{document}


\title{\bf Light dark matter
for Fermi-LAT and CDMS observations}

\author{
Bumseok Kyae$^{(a)}$\footnote{email: bkyae@pusan.ac.kr}
and Jong-Chul Park$^{(b)}$\footnote{email: log1079@gmail.com}
}
\affiliation{$^{(a)}$
Department of Physics, Pusan National University, Busan 609-735, Korea
\\
$^{(b)}$ Department of Physics, Sungkyunkwan University, Suwon 440-746, Korea
}

\begin{abstract}
Light fermionic/scalar dark matter (DM) ($m_{\rm DM}\approx 8~{\rm GeV}$) neutral under the standard model can be responsible for the CDMS and CoGeNT signals, and the Fermi-LAT gamma-ray excesses.
In order to explain them in a relatively simple framework,
we have explored various DM annihilation and scattering processes, discussing important phenomenological constraints coming from particle physics.
Assuming that the two independent observations have a  common DM origin and the processes arise through a common mediator,
DM should annihilate into tau/anti-tau lepton pairs through an $s$-channel,
and scatter with nuclei through a $t$-channel process.
To avoid the $p$-wave suppression, a new Higgs-like scalar field with a mass of ${\mathcal O}$(1) TeV is necessary as a common mediator of both the processes.
We propose a supersymmetric model realizing the scenario.
\end{abstract}


\maketitle


\section{Introduction}

Dark matter (DM) is one of the most important building blocks constituting the universe~\cite{DM-Review}.
According to the recent precise observation from the Planck satellite experiment,
it is believed that DM occupies 27 percent of the present energy density of the universe~\cite{Planck}.
In particular, weakly interacting massive particle (WIMP), which is the most promising DM candidate, is essential for understanding the physics law at the electroweak (EW) scale as well as the structure formation in the universe. Thus, various experiments to explore DM are being carried out on the earth and also outside the atmosphere.


Recently, DM direct detection experiments such as CoGeNT~\cite{CoGeNT}, CDMS-Ge~\cite{CDMS-Ge} and CDMS-Si~\cite{CDMS-Si} have reported the observations of some WIMP-candidate events at $(2-3) \sigma$ confidence level. They are claimed to be interpreted as DM signals with a relatively light mass of $m_{\rm DM} \approx 7-10$ GeV and a spin-independent (SI) elastic scattering cross section per nucleon of $\sigma_{\rm SI} \approx 10^{-41}-10^{-40}\,{\rm cm}^2$.
The best fit point for these three measurements is around $m_{\rm DM} \approx 8$ GeV and $\sigma_{\rm SI} \approx 3\times 10^{-41}\,{\rm cm}^2$. DAMA/LIBRA~\cite{DAMA} and CRESST-II~\cite{CRESST-II} results also support similar parameter regions.
However, all such signals are not exactly compatible with the constraints from XENON10~\cite{XENON10} and XENON100~\cite{XENON100}.
Recently, the authors of Ref.~\cite{Frandsen:2013cna} have pointed out that XENON10's constraint should be weakened, and the XENON10 Collaboration has corrected the old result in the erratum to Ref.~\cite{XENON10}. In addition, the author of Ref.~\cite{Hooper:2013cwa} has studied various uncertainties and assumptions, which could affect XENON100's constraint on light DM.
Very recently, CoGeNT released the updated data, confirming their previous light DM signals \cite{CoGeNT2014}. Under such a tension among the observations, we will particularly focus on the positive results of CDMS and CoGeNT in this paper.\footnote{Right after completion of this work, LUX~\cite{LUX} reported more stringent limit,  constraining all the positive signal regions. In light of LUX, light DM possibilities have been examined in various ways in Refs. \cite{Belanger:2013tla, Gresham:2013mua, Fox:2013pia}.}

If DM annihilates into the standard model (SM) chiral fermions,
it should also emit gamma-rays. Fermi Large Area Telescope (Fermi-LAT)~\cite{Fermi-LAT} is a satellite based experiment measuring cosmic gamma-rays. The recent analyses~\cite{Hooper} based on the data from Fermi-LAT show peaks at energies around $1-10$ GeV in the gamma-ray spectrum coming from around the galactic center.
%
%
It could be interpreted as an evidence of DM annihilation into the leptons $l\overline{l}$ with $m_{\rm DM} \approx 7-12$ GeV or the bottom quarks $b\overline{b}$ with $m_{\rm DM} \approx 25-45$ GeV.
In this case, the required annihilation cross section is $\sigma v \sim 10^{-26}\,{\rm cm}^3/{\rm s}$.\footnote{The current limits on the annihilation of light DM into leptons coming from the cosmic microwave background (CMB) are $\langle \sigma v \rangle_{e\overline{e}} \approx 0.5 - 1 \times 10^{-26}\,{\rm cm}^3/{\rm s}$, $\langle \sigma v \rangle_{\mu\overline{\mu}} \approx 1 - 2 \times 10^{-26}\,{\rm cm}^3/{\rm s}$, and $\langle \sigma v \rangle_{\tau\overline{\tau}} \approx 2 - 3 \times 10^{-26}\,{\rm cm}^3/{\rm s}$ \cite{cmb}. Thus, the case of DM annihilations into $e^-e^+, \mu^-\mu^+, \tau^-\tau^+$ with the same ratio is slightly constrained by the CMB bound. However, if DM mainly annihilates only into $\tau^-\tau^+$, the CMB constraint could be easily avoidable.}


We note that the DM direct detections and the cosmic gamma-ray observation require the similar mass of DM ($m_{\rm DM}\approx 8~{\rm GeV}$) if they all indeed originate from DM.
In this paper, we will discuss the required DM properties and attempt to construct a DM model reflecting them, assuming that the results of Fermi-LAT and CDMS have a common DM origin.
In order to accommodate the two independent classes of experimental results {\it within a single framework}, we will show that

\vspace{0.5cm}

$\bullet$ 8 GeV fermionic/scalar DM, which is assumed to be a SM singlet field,
should annihilate into SM leptons {\it via an $s$-channel}
and scatter with nuclei via a $t$-channel process, and

\vspace{0.5cm}

$\bullet$ both the DM annihilation and scattering processes should be dominantly mediated by a {\it new Higgs-like scalar} field with an $\mathcal{O}$(1) TeV mass.

\vspace{0.5cm}

This paper is organized as follows. In section \ref{sec:DM}, we will discuss the required DM properties, assuming that the Fermi-LAT gamma-ray observation and CDMS DM direct detection have a common DM origin. In section \ref{sec:model}, we will propose a model to satisfy the required conditions discussed in section \ref{sec:DM}.
Section \ref{sec:conclusion} is a conclusion.

%
\section{Dark matter annihilation and scattering}
\label{sec:DM}
%

As mentioned in Introduction, the annihilation process
for the Fermi-LAT observation requires a relatively large cross section ($\sigma v\sim 10^{-26}\,{\rm cm}^3/{\rm s}$).
First, we will discuss the DM annihilation via $s$-channels.
We will assume that the scattering process to explain
the DM direct detection, which is relatively easier to explain, originates from the similar process to that of DM annihilation for the Fermi-LAT observation for simplicity.

\subsection{Annihilation via $s$-channel process}\label{s-channel}

\begin{figure}[t]
\begin{center}
\begin{picture}(300,120)(0,0)

\Photon(20,60)(70,60){4}{4}
\ArrowLine(100,20)(70,60)
\ArrowLine(70,60)(100,100)

\Text(75,0)[]{\bf (a)}

\Text(120,100)[]{\bf $f_{L,R}$}
\Text(125,20)[]{\bf $(f^c)_{R,L}$}
\Text(30,75)[]{\bf $Z_\mu^\prime$}
\Text(90,60)[]{\bf $g^\prime\gamma^\mu$}

\Text(70,85)[]{\bf $\overline{f}_{L,R}$}
\Text(70,35)[]{\bf $f_{L,R}$}


\DashArrowLine(170,60)(220,60){3}
\ArrowLine(250,20)(220,60)
\ArrowLine(250,100)(220,60)

\Text(225,0)[]{\bf (b)}

\Text(265,100)[]{\bf $f_{R}$}
\Text(270,20)[]{\bf $(f^c)_{R}$}
\Text(180,70)[]{\bf $\phi$}
\Text(240,60)[]{\bf $y_f$}

\Text(220,85)[]{\bf $(f^c)_{L}$}
\Text(225,35)[]{\bf $f_{L}$}

\Text(295,60)[]{\bf + ~~h.c.}

\end{picture}
\caption{Vector (a) and scalar mediators (b) coupled to SM chiral fermions. The subscripts $L$ and $R$ indicate the chiralities. $Z_\mu^\prime$ is a new gauge boson. The newly introduced Higgs-like scalar $\phi$ together with the sizable Yukawa coupling $y_f$ is needed for a desired cross section.}\label{fig:Rwing}
\end{center}
\end{figure}
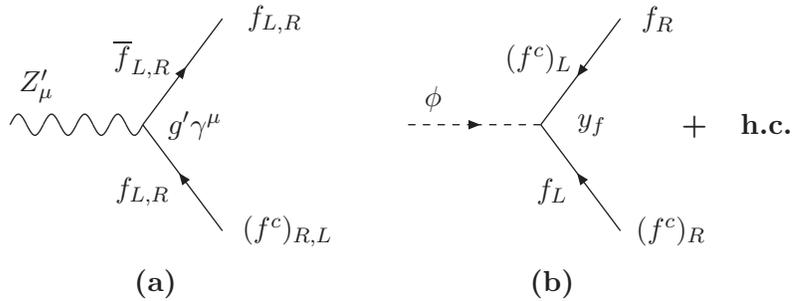

Let us suppose that DM, $X$ and $X^c$ annihilate into SM chiral fermions, $f$ and $f^c$,
\dis{ \label{XXff}
X + X^c \longrightarrow f+ f^c .
}
The masses of $\{X,X^c\}$ are required to be around 8 GeV as mentioned above.
Because of phenomenological and cosmological difficulties, we suppose that 8 GeV DM is not a member of the minimal supersymmetric standard model (MSSM) fields.
In order to pass the EW precision test, we regard them as SM singlets. Since $f$ and $f^c$ are all fermions, a particle mediating the process between $\{X, X^c\}$ and $\{f,f^c\}$  should be a vector or a scalar.
If the chiralities of the final {\it states}, $f$ and $f^c$, are opposite in \eq{XXff}, namely, $\{f_L, (f^c)_R\}$ or $\{f_R, (f^c)_L\}$, only a vector particle or a gauge boson can attach to them as the mediator of DM annihilation.
It is because the relevant vertex in the Lagrangian takes the form of $g^\prime(\overline{f}_{L,R}\gamma^\mu f_{L,R})Z_\mu^\prime$.
See Fig.~\ref{fig:Rwing}-(a).\footnote{
In fact, a scalar mediator also can attach to them, if chirality flipping arises by adding a mass insertion on an external leg in Fig.~\ref{fig:Rwing}-(b). However, $\{f,f^c\}$ are regarded as being quite light in our case, and so such a diagram is suppressed. In this paper, thus, we do not consider such a possibility.
}
In this case, the gauge boson should be an extra gauge boson absent in the SM, because 8 GeV DM $\{X,X^c\}$ cannot carry any SM quantum numbers.\footnote{If our discussion was confined only in DM scattering with nuclei without considering DM annihilation into leptons, an extra gauge boson could also be a possible mediator \cite{Z'medi}.
At one-loop level, the SM gauge fields could also couple to a SM singlet DM \cite{SMmedi}. However, this case turns out to yield too small annihilation cross sections to account for the Fermi-LAT gamma-ray excesses. Throughout this paper, we consider only tree level processes for DM annihilation.
}
SM chiral fermions $\{f,f^c\}$ should also be charged under a new gauge symmetry accompanied with the extra gauge boson. However, the mass of the new gauge boson should be quite heavy ($M_{Z^\prime}\gtrsim 2-3~{\rm TeV}$) to evade the $Z^\prime$ mass constraints by the ATLAS~\cite{ATLAS-Z'} and CMS~\cite{CMS-Z'} Collaborations at the LHC.
If the extra gauge field exclusively couples only to $\tau^{\pm}$ among the leptons, of course, the $Z^\prime$ constraint could be a bit weaker~\cite{ATLAS-Z'-tau}.

%
%
%
%

On the other hand, if the chiralities of the final states, $f$ and $f^c$, are same,
i.e. $\{f_L, (f^c)_L\}$ or $\{f_R, (f^c)_R\}$, the coupling between the SM matter and the mediator should be a type of Yukawa interaction.
In this case, the relevant vertex in the Lagrangian reads
$(f_Lf^c_L)\phi+{\rm h.c.}$ in the Weyl notation
[or $(\overline{f}P_Lf)\phi+{\rm h.c.}$ in the Dirac notation, where $P_L$ denotes the projection operator].
See Fig.~\ref{fig:Rwing}-(b).
Since $f_L$ [$(f^c)_L$] would be an SU(2)$_L$ doublet (singlet) of the SM, the mediator $\phi$ should be a scalar particle carrying the same gauge quantum numbers with the SM Higgs boson. If the mediator is just the SM Higgs \cite{Kim:2008pp}, the coupling $y_f$ must be very small for the SM leptons. Because of this reason, we need to introduce a new Higgs-like scalar $\widetilde L$ with sizable Yukawa couplings in this case. Unlike the SM Higgs, the new SU(2) doublet $\widetilde L$ does not have to get a vacuum expectation value (VEV).

\subsubsection{Fermionic dark matter}

Now let us discuss the spin of $\{X,X^c\}$. If the DM $\{X, X^c\}$ in \eq{XXff} are fermions ($\equiv\{X_F,X_F^c\}$), the mediator linking $\{X_F,X_F^c\}$ and $\{f,f^c\}$ should be a pseudo-scalar or a vector: the annihilation cross section through a real scalar mediator of~\eq{XXff} would be $p$-wave suppressed, unless a fine-tuning effect such as enhancement by a resonance overcomes the suppression.
It is because the initial state, $(X_LX^c_L)$ or $(X_RX^c_R)$, is CP-odd, while the $(X_LX^c_R)$ state is CP-even.
As a result, an initial state $(X_LX^c_L)$ or $(X_RX^c_R)$ [$(X_LX^c_R)$] pair in an $s$-wave state can couple to a pseudo-scalar [vector] mediator.
The needed vertices in the Lagrangian can be provided from $(\overline{X}\gamma^5X){\rm Im}\phi$ in the Dirac notation, which is a part of $(\overline{X}P_LX)\phi +{\rm h.c.}$ (or $X_LX_L^c\phi+{\rm h.c.}$ in the Weyl notation),
and $(\overline{X}\gamma^\mu P_{L,R}X)Z_\mu^\prime$, respectively.
By replacing $\{f,f^c\}$ by $\{X,X^c\}$, the relevant vertices can also be displayed via Fig.~\ref{fig:Rwing}.

For the Majorana DM case ($X_F=X_F^c$), however, the annihilation cross section would be proportional to the mass squared of the final particles, $m_f^2$, if the mediator is a vector field: since the total spin of the initial states, $X_F+X_F$, is zero by the Pauli's exclusion principle in an $s$-wave state, the helicity flipping should arise in Fig.~\ref{fig:Rwing}-(a) such that the chiralities (helicities) of $\{f,f^c\}$ are same (opposite) for the angular momentum conservation.
It is possible by adding a mass insertion on an external leg of $f$ or $f^c$.
Although $\{X_F,X_F^c\}$ exclusively annihilate into $\tau^\pm$, $X_F+X_F\to\tau^++\tau^-$,
the $Z^\prime$ mass bound
is still $1-2$ TeV~\cite{ATLAS-Z'-tau}.
Thus, the suppression factor $(m_\tau/m_{Z'})^2$ is too small to yield the needed annihilation cross section, $\sigma v\sim 10^{-26}\,{\rm cm^3/s}$.


For the Dirac DM case ($X_F\neq X_F^c$)
with a vector field mediation,
it is still hard to get the desired annihilation (and also  scattering) cross section with a gauge boson heavier than $\sim 2-3~{\rm TeV}$, which is required to avoid the $Z^\prime$ constraints~\cite{ATLAS-Z', CMS-Z'} as mentioned above.\footnote{
For the effective gauge couplings  $q_{\rm DM}g^\prime=q_{l}g^\prime=2$ and $q_{q}g^\prime=0.1$, and the masses of $m_{\rm DM}=8~{\rm GeV}$ and $M_{Z^\prime}=1~{\rm TeV}$, one could achieve the desired annihilation and scattering cross sections: $\sigma v\sim 10^{-26}{\rm cm^3/s}$ and $\sigma_{\rm SI} \approx 3\times 10^{-41}\,{\rm cm}^2$. With these couplings, however, the new gauge boson should be heavier than $2-3$ TeV to evade the $Z^\prime$ mass bounds.
}
Of course, an extra gauge field exclusively coupled  only to $\tau^{\pm}$ among the leptons
might loose the experimental bound on $Z^\prime$ mass~\cite{ATLAS-Z'-tau}.
However, a family dependent neutral gauge boson could be a source of a flavor-changing neutral current (FCNC).
To avoid the constraints on the flavor changing muon decay modes, $\mu^-\to e^-e^+e^-$ and $\mu^-\to e^-\gamma$, the breaking scale of such a family-dependent gauge symmetry should be above $3-5$ TeV~\cite{Lfcnc}.
Thus, a heavy gauge boson becomes unavoidable again, which too much suppresses the annihilation cross sections.

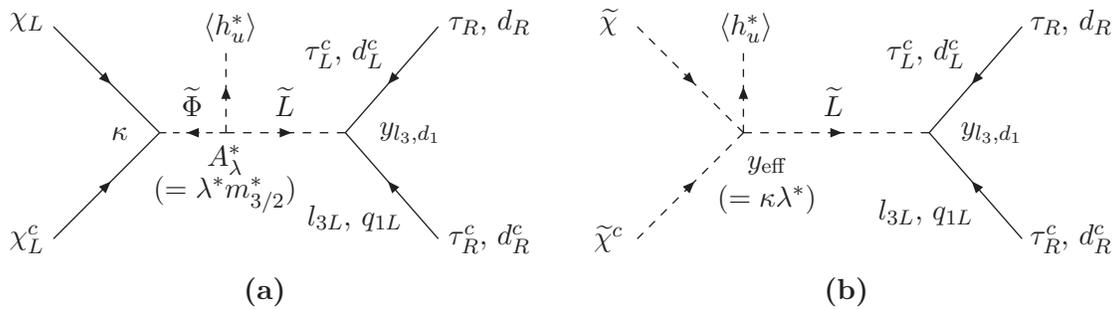
\begin{figure}[t]
\begin{center}
\begin{picture}(420,120)(0,0)

\DashArrowLine(85,60)(130,60){3}
\DashArrowLine(85,60)(60,60){3}
\ArrowLine(165,20)(130,60)
\ArrowLine(165,100)(130,60)

\ArrowLine(20,20)(60,60)
\ArrowLine(20,100)(60,60)
\DashArrowLine(85,60)(85,90){3}

\Text(185,102)[]{{\bf $\tau_R$}, {\bf $d_R$}}
\Text(185,20)[]{{\bf $\tau^c_R$}, {\bf $d^c_R$}}
\Text(72.5,72)[]{\bf $\widetilde{\Phi}$}
\Text(107.5,72)[]{\bf $\widetilde{L}$}
\Text(155,60)[]{\bf $y_{l_3, d_1}$}
\Text(45,60)[]{$\kappa$}
\Text(85,50)[]{$A_\lambda^*$}
\Text(85,37)[]{($=\lambda^* m_{3/2}^*$)}

\Text(130,90)[]{{\bf $\tau^c_L$}, {\bf $d^c_L$}}
\Text(134,28)[]{{\bf $l_{3L}$}, {\bf $q_{1L}$}}


\Text(87,100)[]{\bf $\langle h_u^*\rangle$}
\Text(10,102)[]{\bf $\chi_L$}
\Text(10,20)[]{\bf $\chi_L^c$}

\Text(100,0)[]{\bf (a)}


\DashArrowLine(280,60)(350,60){3}
\ArrowLine(385,20)(350,60)
\ArrowLine(385,100)(350,60)

\DashArrowLine(240,20)(280,60){3}
\DashArrowLine(240,100)(280,60){3}
\DashArrowLine(280,60)(280,90){3}

\Text(405,102)[]{{\bf $\tau_R$}, {\bf $d_R$}}
\Text(405,20)[]{{\bf $\tau^c_R$}, {\bf $d^c_{R}$}}
\Text(315,72)[]{\bf $\widetilde{L}$}
\Text(375,60)[]{\bf $y_{l_3, d_1}$}
\Text(290,48)[]{$y_{\rm eff}$}
\Text(290,35)[]{($=\kappa\lambda^*$)}

\Text(350,90)[]{{\bf $\tau^c_L$}, {\bf $d^c_L$}}
\Text(350,30)[]{{\bf $l_{3L}$}, {\bf $q_{1L}$}}


\Text(282,100)[]{\bf $\langle h_u^*\rangle$}
\Text(230,102)[]{\bf $\widetilde{\chi}$}
\Text(230,20)[]{\bf $\widetilde{\chi}^c$}

\Text(320,0)[]{\bf (b)}

\end{picture}
\caption{{\it $s$-channel annihilation} diagrams of (a) fermionic DM, $\chi +\chi^c\longrightarrow \tau_R +\tau_R^c$,
and (b) scalar DM, $\widetilde{\chi}+\widetilde{\chi}^c\longrightarrow  \tau_R +\tau_R^c$.  The SU(2)$_L$ doublet $\tilde L$ is identified with the scalar $\phi$ in Fig.~\ref{fig:Rwing}(b).
Diagrams (a) and (b) can be understood
as the {\it $t$-channel scattering} diagrams of
$\chi^c_L + d_{L} \longrightarrow \chi^c_R + d_R$
and $\widetilde{\chi}^* + d_{L} \longrightarrow \widetilde{\chi}^* + d_R$, respectively.
In diagram (a), the imaginary (real) parts of $\widetilde{\Phi}$ and $\widetilde{L}$ dominantly mediate the DM annihilation (scattering) process.
$A_\lambda$ denotes the ``A-term'' coefficient in the soft SUSY breaking sector.
}\label{fig:DMannh}
\end{center}
\end{figure}

On the contrary, the desired $s$-channel annihilation cross section of $\{X_F,X_F^c\}$ into SM leptons ($\sigma v\sim 10^{-26}{\rm cm^3/s}$) would be possible,
when this process is mediated by a pseudo-scalar or the imaginary part of a complex scalar:
the part of the real component-mediation would be $p$-wave suppressed.
In this case, the annihilation cross section becomes
proportional to its Yukawa coupling squared with the SM leptons.
As mentioned above, the complex scalar should carry the same gauge quantum numbers with the Higgs in this case.
A possible diagram is displayed in Fig.~\ref{fig:DMannh}-(a), in which we set $X_F=\chi$.
The mediators should be the imaginary parts of $\widetilde{\Phi}$ and $\widetilde{L}$.
For the effective operator of the DM annihilation to be made invariant under the SM gauge group, the Higgs field, $h_d$ or $h_u^*$, should be attached somewhere as an external leg in the diagram as seen in Fig.~\ref{fig:DMannh}-(a).
Note that the process of Fig.~\ref{fig:DMannh}-(a) can be realized in a supersymmetry (SUSY) framework.
In this case, the annihilation cross section is estimated as
\begin{eqnarray}
\langle\sigma v\rangle_{\chi\chi^c \rightarrow l\overline{l}} &=& \frac{\kappa^2 y_l^2}{2\pi}\, M_\chi^2 \left( 1 - \frac{m_l^2}{M_\chi^2} \right)^{1/2}  \left( \frac{A_\lambda v_u/\sqrt{2}}{A_\lambda^2 v_u^2/2 - m_{\widetilde{L}}^2 m_{\widetilde{\Phi}}^2} \right)^2 + ~\mathcal{O}(v^2) \label{fDM-annihilation}\\
&\approx& 1.08 \times 10^{-26}\, {\rm cm}^3/{\rm s}
\nonumber
\end{eqnarray}
for $l=\tau$, $M_\chi=8~{\rm GeV}$, $\kappa=2$, $y_{l_3}=1$, and $A_\lambda = m_{\widetilde{L}} = m_{\widetilde{\Phi}} = 360~{\rm GeV}$.
$v_{u}$ denotes the Higgs VEV: $\langle h_{u, d}^0\rangle = v_{u, d}/\sqrt{2}$ with $v_u^2 + v_d^2 \equiv v^2 \approx (246~{\rm GeV})^2$.\footnote{In all of the analysis of this paper, we use the fixed value of $\tan \beta \equiv v_u/v_d =10$.} This annihilation cross section can be responsible for the gamma-ray excesses of the Fermi-LAT.

%
%
%
%
%
%

So far our discussion has been focused on the $s$-channel annihilation of fermionic DM.
Now we attempt to account for the results of DM direct detection.
In this paper, as mentioned above, we intend to realize it in a common frame:
we assume that the scattering with nuclei in DM direct detection also arises through the similar process, just replacing the SM leptons
by SM quarks, $\{f,f^c\}=\{q,q^c\}$ in Fig.~\ref{fig:DMannh}.
In this case, Fig.~\ref{fig:DMannh}-(a) should be
understood as the $t$-channel diagram representing
\dis{
\chi^c_L + d_{L} \longrightarrow \chi^c_R + d_R \,.
}

In the zero momentum transfer limit, non-vanishing fermionic DM--nucleus elastic scattering cross sections
are allowed only by the following effective operators \cite{micrOMEGAs}:
\dis{
(\overline{X}_F X_F) (\overline{q}q) ~, ~~ (\overline{X}_F \gamma_\mu X_F) (\overline{q}\gamma^\mu q) ~~{\rm [SI]}~, ~~ {\rm and}~~
(\overline{X}_F \gamma_\mu\gamma_5 X_F) (\overline{q}\gamma^\mu\gamma^5 q) ~~ {\rm [SD]}
}
except for tensor operators.
If DM is a scalar ($=X_B$),
\dis{
(X_B^\dagger X_B) (\overline{q}q) ~, ~~{\rm and}~~ (X_B^\dagger \gamma_\mu X_B) (\overline{q}\gamma^\mu q) ~~ {\rm [SI]}
}
also admit non-vanishing scattering cross sections under the zero momentum transfer limit.
Since a vector  mediator has turned out to be undesirable, only $(\overline{X}_F X_F)(\overline{q}q)$ would be promising.
Unlike the annihilation, hence, the real part of a complex scalar should mediate the scattering process
for non-vanishing cross section in the zero momentum transfer limit.
Otherwise, it becomes suppressed.
It implies that once a complex scalar mediator is introduced, the mediation by its real (imaginary) part becomes dominant in the scattering (annihilation) process.
In Fig.~\ref{fig:DMannh}-(a), thus,
the mediation by the real parts of $\widetilde{\Phi}$ and $\widetilde{L}$ survives in the scattering with nuclei.
In this case, the SI scattering cross section per nucleon is
given by~\cite{DM-Review}
\begin{eqnarray}
\sigma_{\rm SI}^{\chi-{\rm nucleon}} = \frac{1}{\pi}\, \frac{M_\chi^2 m_{p,n}^2}{(M_\chi + m_{p,n})^2}\, \frac{1}{A^2}\, [Zf_p + (A-Z)f_n]^2
\end{eqnarray}
with
\begin{eqnarray}
f_{p,n} = \frac{\kappa A_\lambda v_u/\sqrt{2}}{A_\lambda^2 v_u^2/2 - m_{\widetilde{L}}^2 m_{\widetilde{\Phi}}^2}\, \left[ \sum_{q=u,d,s} y_q f_{Tq}^{(p,n)} \frac{m_{p,n}}{m_q} + \frac{2}{27} f_{TQ}^{(p,n)} \sum_{q=c,b,t} y_q \frac{m_{p,n}}{m_q} \right]\,,
\end{eqnarray}
where $A$ and $Z$ denote the atomic mass and proton numbers of the target nuclei, $f_{Tq}^{(p,n)}$ and $f_{TQ}^{(p,n)} \equiv 1 - \sum_{q=u,d,s} f_{Tq}^{(p,n)}$ are the quark form factors in a nucleon state,\footnote{In all the analysis, we use the values given in Ref.~\cite{Belanger:2013oya}.} and $y_{q}$ means the relevant Yukawa coupling.
If we assume that only $y_{d_1}$ is sizable, this cross section is estimated as
\begin{eqnarray}
\sigma_{\rm SI}^{\chi-{\rm nucleon}} \approx 2.84 \times 10^{-41}\, {\rm cm}^2
\label{fDM-scattering}
\end{eqnarray}
for $y_{d_1}=0.013$ and the same parameters with Eq.~(\ref{fDM-annihilation}).
With this Yukawa coupling, fermionic DM also annihilates to $d$-quarks. The annihilation cross section is estimated as $\langle\sigma v\rangle_{\chi\chi^c \rightarrow d\overline{d}} \approx 5.61 \times 10^{-30}\, {\rm cm}^3/{\rm s}$ which is much smaller than current limits on $\langle\sigma v\rangle_{q\overline{q}}$ based on Fermi-LAT gamma-ray observations \cite{Berlin:2013dva}.
If the three down-type quarks have universal Yukawa couplings, $\sigma_{\rm SI}^{\chi-{\rm nucleon}} \approx 3.44 \times 10^{-41}\, {\rm cm}^2$ for the other same parameters with Eq.~(\ref{fDM-scattering}).

\subsubsection{Scalar dark matter}

If DM $\{X, X^c\}$ are scalar fields ($\equiv\{X_B,X_B^c\}$), the mediator should also be a scalar to avoid $p$-wave suppression: the annihilation cross section of $X_B+X_B^c \to f + f^c$ through a vector mediator is $p$-wave suppressed.
%
%
Thus, scalar DM with a scalar mediator is preferred for obtaining the desired values of the annihilation (and also scattering) cross sections,
if DM annihilates dominantly via the $s$-channel.
As discussed above, a Higgs-like scalar $\widetilde L$ is needed again. For invariance of the SM gauge group, the Higgs field should be attached also in the diagram of the effective operator describing~\eq{XXff}. Fig.~\ref{fig:DMannh}-(b) shows one simple possibility, which can also be realized in a SUSY framework.

The annihilation cross section for the scalar DM of
Fig.~\ref{fig:DMannh}-(b) is calculated as
\dis{
&\langle\sigma v\rangle_{\widetilde{\chi}\widetilde{\chi}^c \rightarrow l\overline{l}} = \frac{1}{2\pi}\, \frac{y_{\rm eff}^2 y_l^2 v_u^2}{m_{\widetilde{L}}^4}\, \left( 1 - \frac{m_l^2}{M_{\widetilde{\chi}}^2} \right)^{3/2} + ~\mathcal{O}(v^2) \\
&\approx 1.01 \times 10^{-26}\, {\rm cm}^3/{\rm s}
~\left(\frac{y_{\rm eff}}{0.5} \right)^2
\left(\frac{y_{l_3}}{0.9} \right)^2
\left(\frac{1.2~{\rm TeV}}{m_{\widetilde{L}}} \right)^{4}
\label{sDM-annihilation}
}
for $l=\tau$ and $M_{\widetilde{\chi}}=8~{\rm GeV}$.
%
In addition, the SI scattering cross section for a scalar DM, $\widetilde{\chi}$ in
Fig.~\ref{fig:DMannh}-(b),
$\widetilde{\chi}^* + d_{L} \longrightarrow \widetilde{\chi}^* + d_R$
is given by
\begin{eqnarray}
\sigma_{\rm SI}^{\widetilde{\chi}-{\rm nucleon}} = \frac{1}{4\pi}\, \frac{m_{p,n}^2}{(M_{\widetilde{\chi}}+m_{p,n})^2}\, \frac{1}{A^2}\, [Zf_p+(A-Z)f_n]^2
\end{eqnarray}
with
\begin{eqnarray}
f_{p,n} = \frac{\sqrt{2}y_{\rm eff}v_u}{m_{\widetilde{L}}^2}\, \left[ \sum_{q=u,d,s} y_q f_{Tq}^{(p,n)} \frac{m_{p,n}}{m_q} + \frac{2}{27} f_{TG}^{(p,n)} \sum_{q=c,b,t} y_q \frac{m_{p,n}}{m_q} \right]\,.
\end{eqnarray}
It is estimated as
\begin{eqnarray}
\sigma_{\rm SI}^{\widetilde{\chi}-{\rm nucleon}} \approx 2.94 \times 10^{-41}\, {\rm cm}^2 ~\left(\frac{y_{\rm eff}}{0.5} \right)^2 \left(\frac{y_{d_1}}{0.017} \right)^2 \left(\frac{8~{\rm GeV}}{M_{\widetilde{\chi}}} \right)^{2} \left(\frac{1.2~{\rm TeV}}{m_{\widetilde{L}}} \right)^{4}\,.
\label{sDM-scattering}
\end{eqnarray}
Here we assume that only $y_{d_1}$ is sizable.
With the Yukawa coupling $y_{d_1}=0.017$, scalar DM annihilation cross section to $d$-quarks is estimated as $\langle\sigma v\rangle_{\chi\chi^c \rightarrow d\overline{d}} \approx 1.16 \times 10^{-29}\, {\rm cm}^3/{\rm s}$ which is much less than current limits from Fermi-LAT gamma-ray observations \cite{Berlin:2013dva}.
If the three down-type quarks have universal Yukawa couplings,
the result is altered to $\sigma_{\rm SI}^{\widetilde{\chi}-{\rm nucleon}} \approx 3.57 \times 10^{-41}\, {\rm cm}^2$
for the same parameters with Eq.~(\ref{sDM-scattering}).

\subsection{Annihilation via $t$-channel process}

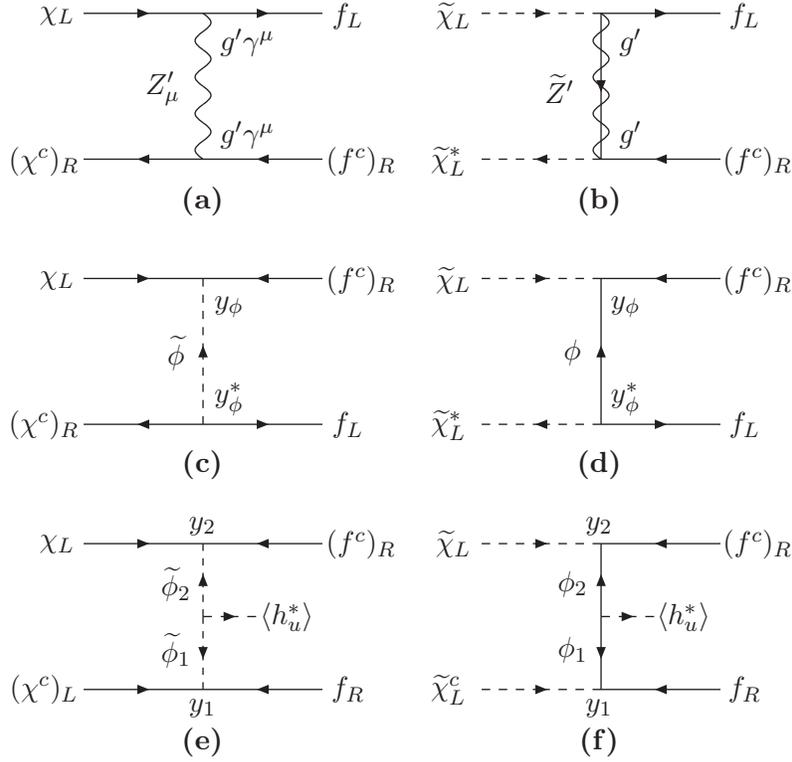
\begin{figure}[t]
\begin{center}
\begin{picture}(300,300)(0,0)

\Photon(75,275)(75,220){3}{4}
\ArrowLine(75,220)(30,220)
\ArrowLine(120,220)(75,220)
\ArrowLine(30,275)(75,275)
\ArrowLine(75,275)(120,275)

\Text(92,265)[]{$g^\prime\gamma^\mu$}
\Text(92,230)[]{$g^\prime\gamma^\mu$}

\Text(75,205)[]{\bf (a)}

\Text(20,275)[]{\bf $\chi_{L}$}
\Text(15,220)[]{\bf $(\chi^c)_{R}$}
\Text(130,275)[]{\bf $f_{L}$}
\Text(135,220)[]{\bf $(f^c)_{R}$}
\Text(60,247.5)[]{\bf $Z_\mu^\prime$}



\Photon(225,275)(225,220){3}{4}
\ArrowLine(225,275)(225,220)

\DashArrowLine(225,220)(180,220){3}
\ArrowLine(270,220)(225,220)
\DashArrowLine(180,275)(225,275){3}
\ArrowLine(225,275)(270,275)

\Text(237,265)[]{$g^\prime$}
\Text(237,230)[]{$g^\prime$}

\Text(225,205)[]{\bf (b)}

\Text(170,275)[]{\bf $\widetilde{\chi}_{L}$}
\Text(168,220)[]{\bf $\widetilde{\chi}_L^*$}
\Text(280,275)[]{\bf $f_{L}$}
\Text(285,220)[]{\bf $(f^c)_{R}$}
\Text(210,247.5)[]{\bf $\widetilde{Z}^\prime$}


\DashArrowLine(75,120)(75,175){3}
\ArrowLine(75,120)(30,120)
\ArrowLine(75,120)(120,120)
\ArrowLine(30,175)(75,175)
\ArrowLine(120,175)(75,175)

\Text(85,165)[]{$y_\phi$}
\Text(85,130)[]{$y_\phi^*$}

\Text(75,105)[]{\bf (c)}

\Text(20,175)[]{\bf $\chi_{L}$}
\Text(15,120)[]{\bf $(\chi^c)_{R}$}
\Text(135,175)[]{\bf $(f^c)_{R}$}
\Text(130,120)[]{\bf $f_{L}$}
\Text(65,147.5)[]{\bf $\widetilde{\phi}$}




\ArrowLine(225,120)(225,175)
\DashArrowLine(225,120)(180,120){3}
\ArrowLine(225,120)(270,120)
\DashArrowLine(180,175)(225,175){3}
\ArrowLine(270,175)(225,175)

\Text(235,165)[]{$y_\phi$}
\Text(235,130)[]{$y_\phi^*$}

\Text(225,105)[]{\bf (d)}

\Text(170,175)[]{\bf $\widetilde{\chi}_{L}$}
\Text(168,120)[]{\bf $\widetilde{\chi}_L^*$}
\Text(285,175)[]{\bf $(f^c)_{R}$}
\Text(280,120)[]{\bf $f_{L}$}
\Text(215,147.5)[]{\bf $\phi$}


\DashArrowLine(75,47.5)(75,20){3}
\DashArrowLine(75,47.5)(75,75){3}

\ArrowLine(30,20)(75,20)
\ArrowLine(120,20)(75,20)
\ArrowLine(30,75)(75,75)
\ArrowLine(120,75)(75,75)
\DashArrowLine(75,47.5)(95,47.5){3}
\Text(107,47.5)[]{\bf $\langle h_u^*\rangle$}

\Text(75,82)[]{$y_2$}
\Text(75,13)[]{$y_1$}

\Text(75,0)[]{\bf (e)}

\Text(20,75)[]{\bf $\chi_{L}$}
\Text(15,20)[]{\bf $(\chi^c)_{L}$}
\Text(135,75)[]{\bf $(f^c)_{R}$}
\Text(130,20)[]{\bf $f_{R}$}
\Text(65,60)[]{\bf $\widetilde\phi_2$}
\Text(65,35)[]{\bf $\widetilde\phi_1$}




\ArrowLine(225,47.5)(225,20)
\ArrowLine(225,47.5)(225,75)
\DashArrowLine(180,20)(225,20){3}
\ArrowLine(270,20)(225,20)
\DashArrowLine(225,47.5)(245,47.5){3}
\Text(257,47.5)[]{\bf $\langle h_u^*\rangle$}
\DashArrowLine(180,75)(225,75){3}
\ArrowLine(270,75)(225,75)

\Text(225,82)[]{$y_2$}
\Text(225,13)[]{$y_1$}

\Text(225,0)[]{\bf (f)}

\Text(170,75)[]{\bf $\widetilde{\chi}_{L}$}
\Text(168,20)[]{\bf $\widetilde{\chi}^c_{L}$}
\Text(285,75)[]{\bf $(f^c)_{R}$}
\Text(280,20)[]{\bf $f_{R}$}

\Text(215,60)[]{\bf $\phi_2$}
\Text(215,35)[]{\bf $\phi_1$}


\end{picture}
\caption{Various $t$-channel processes of DM annihilation into SM leptons. They are $s$-channel processes of DM scattering off SM quarks.
In (a) and (b), $\{Z_\mu^\prime,\widetilde{Z}^\prime\}$ are a new gauge boson and a gaugino absent in the MSSM.
In (c) and (d), the scalar and fermionic mediators, $\{\widetilde{\phi},\phi\}$ carry the opposite SM gauge quantum numbers to $f_L$.
Similarly, in (e) and (f), the gauge quantum numbers of $\{\widetilde{\phi}_{2(1)},\phi_{2(1)}\}$ are opposite to $f_L$ ($f_L^c$).
$\{\widetilde{\phi}_{2(1)},\phi_{2(1)}\}$ are SU(2)$_L$ doublets (singlets).
Unlike the processes in Fig. \ref{fig:DMannh},
the DM annihilation and scattering processes cannot share a common mediator.
}\label{fig:t-channel}
\end{center}
\end{figure}

So far we have discussed $s$-channel DM annihilation and $t$-channel DM scattering.
Now let us explore the possibility of $t$-channel processes for DM annihilation to SM leptons.
In Fig.~\ref{fig:t-channel}, we display various $t$-channel diagrams mediated by vector [$Z_\mu^\prime$], fermion [$\widetilde{Z}^\prime$ and $\phi_{(1,2)}$], and scalar particles [$\widetilde{\phi}_{(1,2)}$].
Since the initial DM states, $\{\chi,\chi^c\}$ or  $\{\widetilde{\chi},\widetilde{\chi}^c\}$ are SM singlets, whereas the final states $\{f,f^c\}$ are SM chiral fermions, the mediators $\{Z_\mu^\prime,\widetilde{Z}^\prime;\widetilde\phi,\phi;\widetilde{\phi}_{1,2},\phi_{1,2}\}$ should carry proper SM gauge charges.
Particularly, in Fig.~\ref{fig:t-channel}-(e) and (f),$\{\widetilde{\phi}_2,\phi_2\}$ [$\{\widetilde{\phi}_1,\phi_1\}$] are SU(2)$_L$ doublets [singlets], because $f_L$ [or $(f^c)_R$]
and $f_L^c$ [or $f_R$] are regarded as an SU(2)$_L$ lepton doublet and a singlet, respectively.
The mediators should be accompanied with their vector-like partners and heavy mass terms.

%
%
%
%

For DM scattering with nuclei, the diagrams in Fig.~\ref{fig:t-channel} should be understood as the diagrams of $s$-channel processes.
In this case, $\{f_L,f_L^c\}$ in Fig.~\ref{fig:t-channel} correspond to SM {\it quark} doublets and singlets, respectively, rather than SM leptons.
Since $\{\widetilde{\phi},\widetilde{\phi}_2;\phi,\phi_2\}$ [$\{\widetilde{\phi}_1;\phi_1\}$] carry the opposite gauge quantum numbers to $f_L$ [$f_L^c$], $s$-channel mediators for DM scattering
should {\it not} be the same as those for the DM annihilation.
That is to say, new mediators with color charges, which should be heavier than $\mathcal{O}(1~{\rm TeV})$ to fulfill the LHC data, are necessary for scattering processes.
As a consequence, the DM scattering and annihilation processes cannot share a common mediator unlike the case in Fig~\ref{fig:DMannh}.
Although this case is not matched to our intention, we will also complete our discussions on it.

For $\{f,f^c\}=\{l,l^c\}$ in Fig.~\ref{fig:t-channel}-(a) and (b), $\{\chi,\chi^c;\widetilde{\chi},\widetilde{\chi}^*\}$ and $\{Z_\mu^\prime,\widetilde{Z}^\prime\}$ can be the right-handed (s)neutrino and e.g. the SU(2)$_R$ gauge field (gaugino) appearing in the  SU(3)$_c\times$SU(2)$_L\times$SU(2)$_R\times$U(1)$_{B-L}$ model, respectively.
Thus, the masses of $\{Z_\mu^\prime,\widetilde{Z}^\prime\}$ are determined by the U(1)$_{B-L}$ breaking scale in such a case.
As mentioned before, they should be heavier than 2--3 TeV to evade the $Z^\prime$ constraint, which too much suppress the annihilation cross section.

In Fig.~\ref{fig:t-channel}-(c) and (d),
the chiralities (helicities)
of produced particles, i.e. a SM chiral fermion and its anti-particle, are opposite (same) as in Fig.~\ref{fig:Rwing}-(a).
For {\it Majorana} DM in Fig.~\ref{fig:t-channel}-(c) and scalar DM in Fig.~\ref{fig:t-channel}-(d),
%
%
the angular momentum of the initial states would be zero.
Hence, the chirality of a produced particle should be flipped with a mass insertion on an external leg such that the angular momentum is conserved in the $s$-wave state.
As a result, the $s$-wave annihilation cross sections are proportional to $m_f^2/m_{\widetilde{\phi}}^4$ and $m_f^2/m_\phi^4$, respectively, and thus suppressed due to the small masses of produced SM leptons.

On the other hand,
the $s$-wave cross section for {\it Dirac} DM in Fig.~\ref{fig:t-channel}-(c) is proportional to $m_\chi^2/m_{\widetilde{\phi}}^4$, which can yield the required annihilation cross section of $\sigma v \approx 10^{-26} {\rm cm}^3/{\rm s}$ with $y_\phi = 1$ and $m_{\widetilde{\phi}} = 160$ GeV.\footnote{The general results on the $t$-channel annihilations can be found e.g. in appendix of Ref.~\cite{scalarDM}.}
However, the mass of $\widetilde{\phi}$ is severely constrained by $leptons + \ptmiss$ signal searches in the LEP~\cite{Slepton-LEP, Slepton-LEP2} and LHC~\cite{Slepton-ATLAS, Slepton-CMS} similar to sleptons, $m_{\widetilde{l}} \gtrsim 320$ GeV. Consequently, even Dirac DM case in Fig.~\ref{fig:t-channel}-(c) cannot provide the desired annihilation cross section.

One could evade $m_f^2/m_{\widetilde{\phi}}^4$ suppression by introducing a coupling of DM to an ${\rm SU(2)}_L$ singlet SM fermion as well as a doublet fermion as in Fig.~\ref{fig:t-channel}-(e).
Note that the chiralities (helicities) of the final states in Fig.~\ref{fig:t-channel}-(e)
are same (opposite) unlike the case in Fig.~\ref{fig:t-channel}-(c) or (d),
since chirality flipping by the Higgs arises in the internal line.
In this case, both an SU(2)$_L$ doublet and a singlet mediator, $\{\widetilde{\phi}_2,\widetilde{\phi}_1\}$ are necessary,
which should be accompanied with their vector-like partners and additional mass terms, as mentioned above.
Even in the case, however, the masses of the mediators turn out to be rather light,  $m_{\widetilde{\phi}_1^l} \approx m_{\widetilde{\phi}_2^l} \approx A_{\phi_1^l \phi_2^l h_u} \approx 240$ GeV with $y_{1, 2}^l = 1$,
to yield the desired annihilation cross section of $\langle\sigma v\rangle_{\chi\chi^c \rightarrow l\overline{l}} \sim 10^{-26} {\rm cm}^3/{\rm s}$.
Here $A_{\phi_1^l \phi_2^l h_u}$ denotes the trilinear coupling of the three scalars, $\widetilde{\phi}_1^l$, $\widetilde{\phi}_2^l$, and $h_u$ in Fig.~\ref{fig:t-channel}-(e).
Such a mass range is already ruled out by the LEP and LHC limits mentioned above.
Allowing rather large couplings of $y_{1, 2}^l = 3$, one can obtain the required annihilation cross section of $\langle\sigma v\rangle_{\chi\chi^c \rightarrow l\overline{l}} \sim 10^{-26} {\rm cm}^3/{\rm s}$ with $m_{\widetilde{\phi}_1^l} \approx m_{\widetilde{\phi}_2^l} \approx A_{\phi_1^l \phi_2^l h_u} \approx 380$ GeV,
which could marginally satisfy the LEP and LHC bounds.
The scattering cross section for Fig.~\ref{fig:t-channel}-(e) is approximated as
\begin{eqnarray}
\sigma_{\rm SI}^{{\chi}-{\rm nucleon}} \sim \frac{660\, y_1^d y_2^d}{\pi}\, \frac{M_\chi^2 m_{p,n}^2}{(M_\chi + m_{p,n})^2} \left( \frac{A_{\phi_1^d \phi_2^d h_u} v_u}{(m_{\widetilde{\phi}_1^d}^2 + m_{\widetilde{\phi}_2^d}^2)^2 - 2 A_{\phi_1^d \phi_2^d h_u}^2 v_u^2} \right)^2 \sim 3 \times 10^{-41}\, {\rm cm}^2
\end{eqnarray}
with $M_\chi = 8~{\rm GeV}$, $y_{1, 2}^d = 1$ and $m_{\widetilde{\phi}_1^d} \approx m_{\widetilde{\phi}_2^d} \approx A_{\phi_1^d \phi_2^d h_u} \approx 1.4$ TeV.
Here we assumed that the nucleonic form factors of the colored mediators are the same as those of the corresponding quarks. Such heavy masses of the colored mediators can evade the LHC constraint.

In the case of scalar DM of Fig.~\ref{fig:t-channel}-(f),\footnote{In this case, all the approximated results, Eqs. (\ref{t-f-annihilation}) and (\ref{t-f-scattering}), are valid when $|m_{\phi_2^{l, d}} - m_{\phi_1^{l, d}}| > y_{\phi_1 \phi_2 h_u}^{l, d} v_u$. In addition, we just assume that $m_{\phi_2^{l, d}} > m_{\phi_1^{l, d}}$ for simplicity.} one can estimate the annihilation cross section as
\begin{eqnarray}\label{t-f-annihilation}
\langle\sigma v\rangle_{\widetilde{\chi}\widetilde{\chi}^c \rightarrow l\overline{l}} \sim \frac{1}{8 \pi} \left( 1 - \frac{m_f^2}{M_{\widetilde{\chi}}^2} \right)^{3/2} \left(\frac{y_1^l y_2^l y_{\phi_1 \phi_2 h_u}^l v_u}{m_{\phi_1^l} m_{\phi_2^l}}\right)^2 \sim 10^{-26}\,  {\rm cm}^3/{\rm s}
\end{eqnarray}
with $M_{\widetilde{\chi}}=8~{\rm GeV}$, $y_{\phi_1 \phi_2 h_u}^l = y_{1, 2}^l = 1$ and $m_{\phi_2^l}/2 \approx m_{\phi_1^l} \approx 900$ GeV, which is heavy enough to satisfy the LEP and LHC's experimental results.
Here $y_{\phi_1 \phi_2 h_u}^l$ denotes the Yukawa coupling of $\phi_1^l$, $\phi_2^l$, and $h_u$ in Fig.~\ref{fig:t-channel}-(f).
For scattering cross section of Fig.~\ref{fig:t-channel}-(f), $\widetilde{\chi}^c_L+d_L^c\rightarrow \widetilde{\chi}_R^*+d^c_R$, we approximately have
\begin{eqnarray}\label{t-f-scattering}
\sigma_{\rm SI}^{\widetilde{\chi}-{\rm nucleon}} \sim \frac{2.6}{\pi}\, \frac{m_{p,n}^2}{(M_{\widetilde{\chi}}+m_{p,n})^2} \left(\frac{y_1^d y_2^d y_{\phi_1 \phi_2 h_u}^d v_u}{m_{\phi_1^d} m_{\phi_2^d}}\right)^2 \sim 3 \times 10^{-41}\, {\rm cm}^2
\end{eqnarray}
with $M_{\widetilde{\chi}}=8~{\rm GeV}$, $y_{\phi_1 \phi_2 h_u}^d =y_{1, 2}^d = 1$ and $m_{\phi_2^d}/2 \approx m_{\phi_1^d} \approx 6.5$ TeV.
Here we assumed again that the nucleonic form factors of the colored mediators are the same as those of the corresponding quarks. Such heavy required masses of the colored mediators can be lowered with smaller couplings, $y_{\phi_1 \phi_2 h_u}^d$ and $y_{1, 2}^d$.

As discussed in this section, the simple DM models with $s$-channel annihilations nicely explain all the experimental observations including CDMS, CoGeNT, and Fermi-LAT signals within a single frame. However, DM in such simple models turns out to overclose the universe ($\Omega_{\rm DM} h^2 \approx 0.25-0.28$). In Section \ref{sec:model}, we will show that there are ways to obtain the correct relic abundance without ruining the salient features of these models.

%
\section{The models}
\label{sec:model}
%

In this section, we attempt to realize the annihilation and scattering processes discussed above by constructing specific models.
We will also briefly discuss how to obtain the correct relic density.

\subsection{Model for $s$-channel annihilation process: Model-I}
\label{subsec:Model-I}

Let us consider the following superpotential:
\dis{ \label{superPot}
W=\left(\kappa\chi\chi^c
+\lambda Lh_u\right)\Phi + y_{l_3} Ll_3e_3^c + y_{d_1}Lq_1d_1^c
+\left(\mu_\chi\chi\chi^c+\mu_LLL^c+\mu_h h_uh_d +\frac{M^{ij}}{2}\nu_i^c\nu_j^c \right) ,
}
where $\kappa$, $\lambda$, and $y_{l_i, d_i}$ denote dimensionless coupling, while $\mu_{\chi}$, $\mu_L$, $\mu_h$, and $M^{ij}$ are dimensionful parameters breaking the global U(1)$_{PQ}$ symmetry.
We suppose that they are all real parameters for simplicity.
Note that as seen in Eqs.~(\ref{fDM-annihilation}), (\ref{fDM-scattering}) and (\ref{sDM-annihilation}), (\ref{sDM-scattering}), even if $y_{l_3}$ should be much larger than $y_{d_1}$, it does not affect the DM scattering process with nuclei.
The global quantum numbers of the superfields in \eq{superPot} are listed in Table \ref{tab:Qnumb1}.
\begin{table}[!h]
\begin{center}
\begin{tabular}
{c|ccccc|ccccc}
{\rm Superfields}  &   ~~$\chi$ ~ & ~$\chi^c$~   & ~$\Phi$~
& ~$L$~  & ~$L^c$~  &  ~$h_d$~ & ~$h_u$~  & ~$q_i$, $l_i$~ & ~$d_i^c$, $e_i^c$~ & ~$u_i^c$, $\nu_i^c$~
  \\
\hline
 U(1)$_{PQ}$  & $-1$ & $-1$ & $2$ & ~$0$~ & $2$ & $0$ & $-2$ & ~$\frac32$ & $-\frac32$ & ~$\frac12$
\end{tabular}
\end{center}\caption{
U(1)$_{PQ}$ charge assignment for the extra and MSSM superfields in Model-I. Dark matter $\{\chi,\chi^c\}$ are SM singlets, but can carry gauge charges under a hidden gauge group. $\Phi$ is a SM singlet. $\{L,L^c\}$ are vector-like leptons under the SM gauge group. The other superfields are those of the MSSM.
}\label{tab:Qnumb1}
\end{table}
In \eq{superPot}, we dropped the ordinary Yukawa interaction terms in the MSSM, which are consistent with the U(1)$_{PQ}$ symmetry.
In principle, the superfield $L$, which carries the same gauge quantum numbers with the MSSM Higgs $h_d$,
can couple to the three families of the MSSM matter superfields, $\{l_{1,2,3}, e^c_{1,2,3}; q_{1,2,3}, d^c_{1,2,3}\}$.
However, we assume that except the couplings $y_{l_3}$ and $y_{d_1}$ in \eq{superPot},
all other Yukawa couplings of $L$ to the MSSM matter are quite suppressed.

In contrast to $L$, the superfield $L^c$, which carries the same gauge quantum numbers with $h_u$ but the opposite global charge to it,
does not couple to the MSSM matter at the renormalizable level due to U(1)$_{PQ}$.
In fact, $\{L,L^c\}$ could induce FCNC, if $L^c$ also had a sizable coupling to the $u$-type quarks and right-handed neutrinos. We assume that the mixings in the CKM and PMNS matrices mainly originate from the $u$-type quark and neutrino sectors, respectively. Only if the Yukawa couplings of the $d$-type quark and charged lepton sectors are approximately block-diagonal, thus, the unwanted FCNC could be avoided.

We note that U(1)$_{PQ}$ disallows the $R$-parity violating couplings and also the terms leading to dimension five proton decays, $q_iq_jq_kl_l$ and $u^c_iu^c_jd^c_ke^c_l$ in the superpotential.
We suppose that U(1)$_{PQ}$ is broken by the VEVs of two singlets carrying $+1$ and $-1$ U(1)$_{PQ}$ charges, $\langle P\rangle$ and $\langle Q\rangle$, respectively. They are assumed to be of order $10^{10}~{\rm GeV}$. By the VEVs, U(1)$_{PQ}$ is broken to $Z_2$, which can be identified with the matter parity in the MSSM. Then, $\mu_{\chi,h}$ can be replaced by $\rho_{\chi,h}\langle P\rangle^2/M_P$, while $\mu_L$ and $M^{ij}$ by $\rho_{L}\langle Q\rangle^2/M_P$ and $\rho^{ij}\langle Q\rangle$, respectively.
Here the $\rho_i$ parameters denote dimensionless couplings, and $M_P$ ($\approx 2.4\times 10^{18}~{\rm GeV}$) is the reduced Planck mass.
Thus, the $\mu$ parameters in \eq{superPot} are of order EW  or TeV scale, while $M_{ij}$ of order $10^{10}~{\rm GeV}$. By the VEVs, the unwanted terms $q_iq_jq_kl_l$ and $u^c_iu^c_jd^c_ke^c_l$ can be induced, but they are extremely suppressed by factors of order $\langle Q\rangle^6/M_P^7$ ($\sim 10^{-48}/M_P$) and $\langle P\rangle^2/M_P^3$ ($\sim 10^{-16}/M_P$), respectively. Thus, they cannot induce observable proton decays.

We note that the odd parity of an {\it additional} $Z_2$ symmetry can be assigned to $\{\chi,\chi^c\}$. Hence,  both the bosonic and fermionic modes of $\{\chi,\chi^c\}$ could remain stable and be DM components in principle.
%
%
The fermionic modes of them compose a Dirac spinor.
%
%
Provided that the scalar components of $\{\chi,\chi^c\}$ are quite heavier than the fermionic ones,
they can decay to the fermionic components, however,  $\widetilde{\chi}^{(c)}\to \chi^{(c)}+\Phi$, if kinematically allowed.
In this case, only the fermionic components of $\{\chi,\chi^c\}$ can be DM.
Through the mediation of the imaginary (real) parts of $\{\widetilde{\Phi},\widetilde{L}\}$
and the mixing effect by SUSY breaking $A$-term corresponding the $\lambda$ term in \eq{superPot},
the fermionic DM can annihilate into $\tau,\tau^c$ (scatter with the d-quark).
See Fig.~\ref{fig:DMannh}-(a).
In this case, we should assume that the dimensionless coupling $\rho_\chi$ is relatively small [$\sim {\cal O}(0.1)$] for 8 GeV DM mass.

On the other hand, if the fermionic components are heavier enough than the bosonic ones, they can decay to the bosonic components, $\chi^{(c)}\to \widetilde{\chi}^{(c)}+\Phi$. In this case, only the bosonic modes of $\{\chi,\chi^c\}$, i.e. $\{\widetilde \chi, \widetilde \chi^c\}$, can be DM components. They can also annihilate into the fermionic modes of $\{\tau,\tau^c\}$ (scatter with the $d$-quark) through the mediation of the scalar component of $L$ ($\equiv \widetilde L$). See Fig.~\ref{fig:DMannh}-(b). The left vertex in Fig.~\ref{fig:DMannh}-(b) is given by the cross term of $|\partial W/\partial \Phi|^2$. Hence, the effective coupling in Fig.~\ref{fig:DMannh}-(b) is given by $\kappa\lambda^*$ in this model.
It can be also provided by the $A$-term of Fig.~\ref{fig:DMannh}-(a).

Actually, the required mass of $\{\widetilde \chi, \widetilde \chi^c\}$, $m_{\rm DM}\sim 8$ GeV is quite small as an elementary scalar mass. One simple way to obtain such a small mass is just to assume them as pseudo-Goldstones, which can be  remnants after spontaneous breaking of a large gauge group in the hidden sector. Alternatively, one can assume that their soft mass squared ($\equiv m^2_{\widetilde{\chi}}$) is negative, and their physical mass ($M_{\widetilde{\chi}} = \sqrt{m^2_{\widetilde{\chi}} + \mu_\chi^2}$) is properly tuned between the negative soft mass squared and the SUSY mass squared $\mu_\chi^2$.
Then, the bosonic components become lighter than the fermionic ones. It is possible through the renormalization group effects, if the Yukawa coupling $\kappa$ is of order unity and $m^2_{\widetilde{\chi}}$ is positive and not excessively large at the UV cutoff scale. The Landau-pole problem associated with a relatively large $\kappa$ can be avoided by assigning hidden gauge charges to the superfields $\{\chi,\chi^c\}$.
Then, the hidden gauge group embeds the $Z_2$ symmetry, under which $\{\chi,\chi^c\}$ carry the odd parity.

Only with the models discussed in Section~\ref{s-channel} or even with the superpotential \eq{superPot}, DM would not provide the correct relic density
($\Omega_{\rm DM} h^2 \approx 0.25-0.28$), which thus needs to be reduced.
%
%
In order to open a new annihilation channel of $\{\chi,\chi^c\}$ or $\{\tilde\chi,\tilde{\chi}^c\}$,
one could introduce e.g. light singlets $\{\Psi,\Psi^c\}$
 ($m_\Psi,m_{\Psi^c}\ll m_\chi,m_{\chi^c}$), and a term $W\supset \kappa^\prime \Psi\Psi^c\Phi$ in the superpotential, where $\kappa^\prime$ is a coupling constant.
We assume that the fermionic modes of $\{\Psi, \Psi^c\}$ are light enough.
Then, $\{\chi,\chi^c\}$ or $\{\tilde\chi,\tilde{\chi}^c\}$ could be in thermal equilibrium state with $\{\Psi, \Psi^c\}$,
$\chi\chi^c\longleftrightarrow \Psi\Psi^c$ or
$\tilde{\chi}\tilde{\chi}^c\longleftrightarrow \Psi\Psi^c$ via $\tilde\Phi$ in the early universe.
$\{\Psi,\Psi^c\}$ could be in the thermal bath through
$\Psi\Psi^c\longleftrightarrow\tau \tau^c,  d d^c$.
This diagram can be drawn by replacing $\{\chi,\chi^c\}$ by $\{\Psi,\Psi^c\}$ in Fig.~\ref{fig:DMannh}-(a).
$\{\chi,\chi^c\}$ or $\{\tilde\chi,\tilde{\chi}^c\}$ decouple first
from the thermal bath because of their relatively heavier masses and become dominant thermal relic.
$\{\Psi,\Psi^c\}$ decouple later from the thermal bath, and so their number density can be assumed to be small enough.
Since they are relatively light,
we can have an additional annihilation channel
$\tilde\chi\tilde{\chi}^c \longrightarrow \Psi\Psi^c$ at the present.

\subsection{Model for $t$-channel annihilation process: Model-II}

As discussed above, the desired $t$-channel annihilation and $s$-channel scattering cross sections can be obtained also from the processes in Fig.~\ref{fig:t-channel}-(e) and (f).
However, these cases are not possible only with a common mediator.
Nonetheless, we will briefly present a model for
Fig.~\ref{fig:t-channel}-(f).
DM can be a scalar field rather than a fermion
by the mechanism explained in section \ref{subsec:Model-I}.
Moreover, if the masses of the mediators are given by the values required in Fig.~\ref{fig:t-channel}-(f), the processes of Fig.~\ref{fig:t-channel}-(e) become  relatively suppressed.

It can be realized from the following superpotential:
\dis{ \label{superPot2}
W=y_{2} D_H^c \psi_D\chi + \mu_DD_HD_H^c
+ \left(y_{1} S_H  \psi_S^c\chi^c
+ y_h D_H^c S_H h_u + \mu_SS_HS_H^c \right)  ,
}
where $\psi_D$ and $\psi_S$ denote the MSSM SU(2)$_L$ doublets and singlets. For simplicity and avoiding FCNC, we again assume that $y_{2}$ and $y_{1}$ are sizable
only for one generation of quarks and leptons,
$\psi_D=l_3,~q_1$ and $\psi_S^c=\tau^c,~d^c$.
$\{D_H,D_H^c\}$ and $\{S_H^c,S_H\}$ are extra vector-like doublets and singlets.
The gauge quantum numbers of $D_H$ ($D_H^c$)
and $S_H^c$ ($S_H$) are the same as (opposite to)
$\psi_D$ and $\psi_S^c$, respectively.
$\{D_H,D_H^c\}$ and $\{S_H,S_H^c\}$ get Dirac masses from the $\mu$ parameters in \eq{superPot2},
which break the U(1)$_{PQ}$ symmetry.
As in the previous model, they can be replaced by
VEVs of the spurion superfields, $\langle P\rangle^2/M_P$ and $\langle Q\rangle^2/M_P$.
The global charge assignment of the U(1)$_{PQ}$ for $\{D_H,D_H^c\}$ and $\{S_H,S_H^c\}$ is listed in Table~\ref{tab:Qnumb2}.
For the U(1)$_{PQ}$ charges of $\{\chi,\chi^c\}$ and the MSSM superfields, see Table~\ref{tab:Qnumb1}.
$f_L$ ($f^c_L$) in Fig.~\ref{fig:t-channel} can be  identified with $\psi_D$ ($\psi_S^c$).
Then, $\{\phi_1,\phi_2\}$ are regarded as $\{S_H,D_H^c\}$.

\begin{table}[!h]
\begin{center}
\begin{tabular}
{c|cccc}
{\rm Superfields}  &   ~~$D_H$ ~ & ~~$D_H^c$~   & ~~$S_H$~
& ~~$S_H^c$~
  \\
\hline
 U(1)$_{PQ}$  & $-\frac32$ & $-\frac12$ & ~$\frac52$ & $-\frac12$
\end{tabular}
\end{center}\caption{
U(1)$_{PQ}$ charge assignment for the extra vector-like superfields in Model-II. While $D_H$ ($D_H^c$) has the same (opposite) gauge quantum numbers with $l_3$ or $q_1$,
the gauge quantum numbers of
$S_H^c$ ($S_H$) are the same with (opposite to) $\tau^c$ or $d^c$.
}\label{tab:Qnumb2}
\end{table}

For Fig.~\ref{fig:t-channel}-(c) and (d),
only the first two terms in \eq{superPot2} are enough.
As discussed above, however, the resulting annihilation cross section is too small. Thus, the other terms in \eq{superPot2} are also needed for Fig.~\ref{fig:t-channel}-(f).
%

%
\section{Conclusion}
\label{sec:conclusion}

We have discussed various possibilities of DM annihilation and scattering processes to explain the Fermi-LAT's observations and CDMS experiments in a common framework. If 8 GeV fermionic/scalar DM, which is assumed to be a SM singlet, annihilates into tau/anti-tau pair via an $s$-channel and scatters off nuclei via a $t$-channel process, the desired cross sections can be achievable
with a common mediator, avoiding $p$-wave suppression.
The mediator should be a scalar field carrying the same gauge quantum number with the SM Higgs boson, but with a mass of $\mathcal{O}$(1) TeV for fermionic or scalar DM.
Only with a simple model, however, DM overcloses the universe. We have proposed a SUSY model realizing the desired features.
By extending the model, the correct DM density can also be addressed in this framework.

%

\section*{Acknowledgment}

This work is supported by Basic Science Research Program through the
National Research Foundation of Korea (NRF) funded by the Ministry of Education, Grant No. 2013R1A1A2006904 (B.K.),
2011-0029758, and 2013R1A1A2061561 (J.-C.P.),
and also in part
by Korea Institute for Advanced Study (KIAS) grant funded by the Korea government (B.K.).



\end{document}